\documentclass[secnumarabic,amssymb,nobibnotes,aps,prd]{revtex4}
\usepackage{amssymb, amsmath}
\usepackage{mathbbol}
\usepackage{graphicx}
\usepackage{dcolumn}
\usepackage{bm}
\usepackage{hyperref}
\usepackage{color}
\begin{document}

\title{ Relativistic Correction to Charmonium Dissociation Temperature }
\author{Xingyu Guo}
\author{Shuzhe Shi}
\author{Pengfei Zhuang}
\affiliation{Physics Department, Tsinghua University, Beijing
100084, China }
\date{\today}

\begin{abstract}
By solving the covariant relativistic Schr\"odinger equations for a
pair of heavy quarks, we obtained the wave functions for the ground
and excited quarkonium states at finite temperature. In comparison
with the non-relativistic calculation, the $J/\psi$ dissociation
temperature determined by the infinity size or zero binding energy
of the system increases $7\%-13\%$, when the central potential
varies between the free energy and internal energy.
\end{abstract}

\maketitle

It is widely accepted that there exists a quantum chromodynamics
(QCD) phase transition from hadron gas to a new state of matter, the
quark-gluon plasma (QGP) at finite temperature and baryon density.
To probe the realization of such a phase transition in relativistic
heavy ion collisions, some signatures of the new state of matter
have been discussed for decades\cite{qgp}, and among which the
quarkonium suppression is considered as a smoking gun of the
formation of QGP\cite{satz1}.

A quarkonium is a deeply bound state of a pair of heavy quarks, its
dissociation temperature $T_D$ in a hot medium should be higher than
the critical temperature $T_c$ for the deconfinement of light
hadrons. Since a quarkonium is so heavy, one normally uses
non-relativistic Schr\"odinger equation to describe its dynamical
evolution at finite temperature. The free energy $F$ between a pair
of heavy quarks can be extracted from the lattice QCD
simulations\cite{digal,kaczmarek}. Taking $F$ and the internal energy
$U$ as the two limits of the heavy quark potential $V$, which
correspond respectively to a slow and a rapid quarkonium
dissociation in the hot medium, the $J/\psi$ dissociation
temperature $T_D$ determined by the infinity size or the zero
binding energy of the system is in between $1.2 T_c$ and $2
T_c$\cite{digal,satz2}.

A nature question we ask ourselves is the relativistic correction to
the dynamical evolution of a quarkonium in the hot medium. The
correction to a bottonium is expected to be neglected safely, but
for a lighter quarkonium like $J/\psi$, the correction might be
remarkable. The two-body Dirac equation (TBDE)\cite{t1,t2,t3,t4,t5,long} of
constrained dynamics was successfully applied to the relativistic
description of light meson spectra\cite{s1,s2,s3,s4,crater2,crater1} in vacuum. In this
Letter we take the TBDE to calculate the charmonium wave functions
at finite temperature and see the relativistic correction to the
dissociation temperature.

We can qualitatively estimate the relativistic effect on the
quarkonium dissociation before a strict calculation. Neglecting the
quark spin, the relative part of the Hamiltonian for a pair of heavy
quarks can be approximately written as a non-relativistic form
\begin{equation}
\label{estimate} H=\sqrt{\mu^2+p^2}-\mu+V(r)\simeq {p^2\over
2\mu}+V_{eff}
\end{equation}
with an effective potential
\begin{equation}
\label{veff}
V_{eff}=V-{p^4\over 8\mu^3},
\end{equation}
where $\mu=m_Q/2$ is the reduced mass with $m_Q$ being the heavy
quark mass. Since the relativistic correction leads to a deeper
potential well, $V_{eff}< V$, the quarkonium becomes a more deeply
bound state and the temperature needed to dissociate the quarkonium
should be higher.

We now calculate the wave functions for a pair of heavy quarks in
the frame of TBDE. Taking Pauli reduction and scale
transformation\cite{333}, the Dirac equation can be expressed as a
covariant relativistic Schr\"odinger equation for a four-component
spinor\cite{long}. Explicitly, the radial motion relative to the
center of mass is controlled by the following four
equations\cite{crater2},
\begin{eqnarray}
\label{wave3}
&&\left[-{d^2\over dr^2}+{J(J+1)\over
r^2}+2m_wB+B^2-A^2+2\epsilon_wA+\Phi_D-2\Phi_{SO}+\Phi_{SS}+2\Phi_T-2\Phi_{SOT}\right]u_1^0=b^2 u_1^0,\nonumber\\
&&\left[-{d^2\over dr^2}+{J(J-1)\over
r^2}+2m_wB+B^2-A^2+2\epsilon_wA+\Phi_D+2(J-1)\Phi_{SO}+\Phi_{SS}+{2(J-1)\over 2J+1}(\Phi_{SOT}-\Phi_T)\right]u_1^+\nonumber\\
&+&{2\sqrt{J(J+1)}\over
2J+1}\left(3\Phi_T-2(J+2)\Phi_{SOT}\right)u_1^-=b^2 u_1^+,\nonumber\\
&&\left[-{d^2\over dr^2}+{(J+1)(J+2)\over
r^2}+2m_wB+B^2-A^2+2\epsilon_wA+\Phi_D-2(J+2)\Phi_{SO}+\Phi_{SS}+{2(J+2)\over 2J+1}\left(\Phi_{SOT}-\Phi_T\right)\right]u_1^-\nonumber\\
&+& {2\sqrt{J(J+1)}\over
2J+1}\left(3\Phi_T+2(J-1)\Phi_{SOT}\right)u_1^+=b^2 u_1^-
\end{eqnarray}
for the spin triplet $u_1^0, u_1^+$ and $u_1^-$ with quantum numbers
$n^{2s+1}L_J=n^3L_L,\ n^3L_{L+1}$ and $n^3L_{L-1}$, and
\begin{equation}
\label{wave1}
\left[-{d^2\over dr^2}+{J(J+1)\over
r^2}+2m_wB+B^2-A^2+2\epsilon_wA+\Phi_D-3\Phi_{SS}\right]u_0=b^2 u_0
\end{equation}
for the spin singlet $u_0$ with quantum numbers $n^1L_L$, where
$b^2=(m_m^2-4m_Q^2)/4$ is the energy eigenvalue in the meson rest
frame with $m_m$ being the meson mass, $n$ is the principal quantum
number, and $L, s$ and $J$ are the orbital, spin and total angular
momentum numbers. Note that the components $u_1^+$ with $J=L+1$ and
$u_1^-$ with $J=L-1$ are coupled to each other. Following the
notations in \cite{crater2}, we have separated the central potential
into two parts,
\begin{equation}
\label{vr} V(r)=A(r) + B(r),
\end{equation}
and the abbreviations for the Darwin, spin-spin, spin-orbit and
tensor terms introduced in the dynamical equations are defined as
\cite{crater2}
\begin{eqnarray}
\label{abbre} &\Phi_D& = M+F'^2+K'^2-\nabla^2F+2K'P-2\left(F'+{1\over r}\right)Q,\nonumber\\
&\Phi_{T}&= {1\over 3}\left[N+2F'K'-\nabla^2K+\left(3F'-K'+{3\over r}\right)P+ \left(F'-3K'+{1\over r}\right)Q\right]\nonumber\\
&\Phi_{SO}&=-{F'\over r}+K'P-\left(F'+{1\over r}\right)Q,\nonumber\\
&\Phi_{SS}&=O+{2\over 3}F'K'-{1\over 3}\nabla^2K+{2\over 3}K'P-2\left(F'+{1\over 3r}\right)Q,\nonumber\\
&\Phi_{SOT}&=-{K'\over r}+\left(F'+{1\over r}\right)P-K'Q,\nonumber\\
&F&={1\over 2}L-{3\over 2}G,\nonumber\\
&G&=-{1\over 2}\ln \left(1-2{A\over m_m}\right),\nonumber\\
&K&={1\over 2}L+{1\over 2}G,\nonumber\\
&L&=\ln\sqrt{1+{2m_wB+B^2\over m_Q^2\left(1-2A/m_m\right)}},\nonumber\\
&M&=-{1\over 2}\nabla^2 G+{3\over 4}G'^2+G'F'-K'^2,\nonumber\\
&N&={1\over 3}\left[\nabla^2K-{1\over 2}\nabla^2G+{3\over
2}{G'-2K'\over r}+F'(G'-2K')\right]\nonumber\\
&O&={1\over 3}\nabla^2(K+G)-{1\over 3}F'(G'+K')-{1\over
2}G'^2,\nonumber\\
&P&={\sinh 2K\over r},\nonumber\\
&Q&={\cosh 2K-1\over r}
\end{eqnarray}
and $m_w=m_Q^2/m_m, \epsilon_w=(m_m^2-2m_Q^2)/(2m_m), F'=dF/dr,
G'=dG/dr$ and $K'=dK/dr$.

We now focus on the charmonium states. The $J/\psi$ state includes
two components $1^3S_1$ and $1^3D_1$ determined simultaneously by
the two coupled equations of (\ref{wave3}) with energy eigenvalue
$b^2=(m_{J/\psi}^2-4m_c^2)/4$. Similarly, the two components $
2^3S_1$ and $2^3D_1$ for the state $\psi'$ are controlled by the two
coupled equations but with eigenvalue $b^2=(m_{\psi'}^2-4m_c^2)/4$.
The $\chi_0$ state $1^3P_0$ is described by the last equation of
(\ref{wave3}) with disappeared component $u_1^+$, and the $\chi_1$
state $1^3P_1$ is characterized by the independent equation
(\ref{wave1}) for $u_1^0$.

In vacuum, the potential between two quarks is usually taken as the
Cornell form, including a Coulomb-like part which dominates the wave
functions around $r=0$ and a linear part which leads to the quark
confinement,
\begin{eqnarray}
\label{vacuum}
&A(r)&=-{\alpha\over r},\nonumber\\
&B(r)&=\sigma r.
\end{eqnarray}
The three parameters in the model, namely the charm quark mass $m_c$
in the Schr\"odinger equation and the two coupling constants
$\alpha$ and $\sigma$ in the potential, can be fixed by fitting the
charmonium masses in vacuum.  By taking $m_c=1.422$ GeV,
$\alpha=0.492$ and $\sigma=0.186$ GeV$^2$, we obtain the charmonium
masses $m_{J/\psi}=3.113$ GeV, $m_{\psi'}=3.692$ GeV,
$m_{\chi_0}=3.404$ GeV and $m_{\chi_1}=3.504$ GeV, which are very
close to the experimental values $m_{J/\psi}=3.097$ GeV,
$m_{\psi'}=3.686$ GeV, $m_{\chi_0}=3.415$ GeV and $m_{\chi_1}=3.511$
GeV.

The charmonium dissociation temperature in a hot medium can be
determined by solving the corresponding dynamical equation for the
$c \bar c$ system with potential $V$ between the two heavy quarks at
finite temperature. The potential depends on the dissociation
process in the medium. At the moment, we know only its two limits.
For a rapid dissociation where there is no heat exchange between the
heavy quarks and the medium, the potential is just the internal
energy $U$, while for a slow dissociation, there is enough time for
the heavy quarks to exchange heat with the medium, the free energy
$F$ which can be extracted from the lattice calculations is taken as
the potential\cite{shuryak}. From the thermodynamic
relation $F = U-T {\cal S}$ where ${\cal S}$ is the entropy,
the potential well is deeper for $V=U$ and therefore the
dissociation temperature of charmonium states with potential $V =U$
is higher than that with $V = F$. In the literatures, a number of
effective potentials in between $F$ and $U$ have been used to
evaluate the charmonium evolution in QCD
medium\cite{digal,satz2,shuryak,wong}. In non-relativistic case, H.Satz and his
collaborators\cite{digal,satz2} solved the Schr\"odinger equation and found the
dissociation temperatures $T_d/T_c = 2.1$ and $1.26$ for $V=U$ and
$V=F$ respectively, where $T_c = 165$ MeV\cite{digal,satz2} is the
critical temperature for the deconfinement.

Considering the Debye screening at finite temperature, the potential
$V=F=A+B$ can be written as\cite{digal,satz2}
\begin{eqnarray}
\label{vf}
A(r,T)&=&-{\alpha\over r} e^{-\mu r},\nonumber\\
B(r,T)&=&{\sigma\over \mu}\left[{\Gamma\left({1\over 4}\right)\over
2^{3\over 2}\Gamma\left({3\over 4}\right)} -{\sqrt{\mu r}\over
2^{3\over 4}\Gamma\left({3\over 4}\right)} K_{1\over 4}\left(\mu^2
r^2\right) \right]-\alpha\mu,
\end{eqnarray}
where $\Gamma$ is the Gamma function, $K$ is the modified Bessel
function of the second kind, and the temperature dependent parameter
$\mu(T)$, namely the screening mass or the inverse screening radius,
can be extracted from fitting the lattice simulated free
energy~\cite{digal,kaczmarek}. From the known free energy $F$,
one can then obtain the other limit of the potential, $V=U=F+T {\cal
S}$ by taking ${\cal S}=-\partial F/\partial T$.

We use the inverse power method\cite{crater3} to solve the
Schr\"odinger equations (\ref{wave3}) and (\ref{wave1}) and obtain
the radial wave functions
\begin{equation}
\label{radial}
\Psi(r,T)={u(r,T)\over r}
\end{equation}
for the charmonium states $J/\psi$, $\psi'$ and $\chi_c$ at finite
temperature.

\begin{figure}[!h]
\begin{center}
\includegraphics[width=0.35\textwidth]{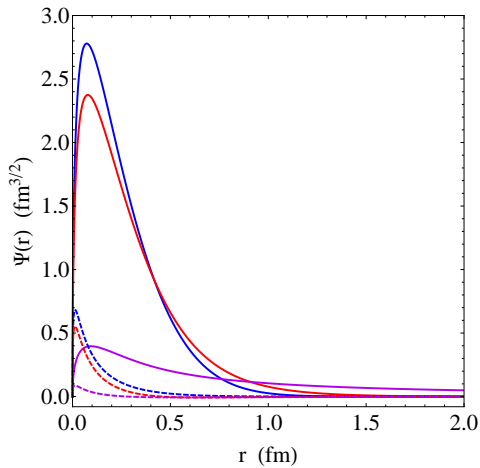}
\caption{ The $S$-wave (solid lines) and $D$-wave (dashed lines)
functions for the $J/\psi$ meson at three temperatures in the limit
of quark potential $V=F$. From the top down the temperature $T$ is
zero, critical temperature $T_c$ and relativistic dissociation
temperature $T_d$.} \label{fig1}
\end{center}
\end{figure}
\begin{figure}[!h]
\begin{center}
\includegraphics[width=0.35\textwidth]{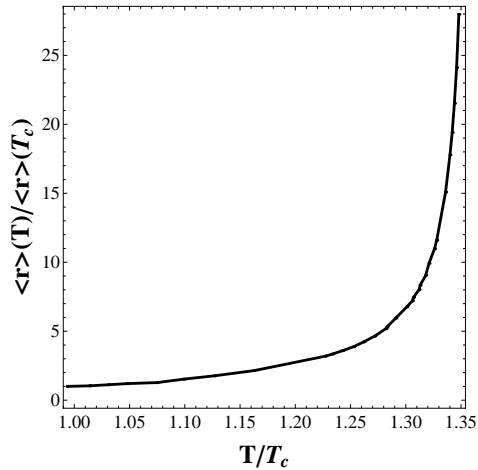}
\caption{ The average size of the $J/\psi$ meson in the quark-gluon
plasma in the limit of potential $V=F$. The temperature and average
size are scaled by their corresponding values at the critical point
of the deconfinement. } \label{fig2}
\end{center}
\end{figure}

The $S-$ and $D-$wave functions for the meson $J/\psi$ at different
temperature are shown in Fig.\ref{fig1} in the limit of $V=F$. The
wave functions in vacuum are very similar to the results obtained in
\cite{crater2}, and the component $1^3S_1$ dominates the state at any
temperature. With increasing temperature, both the $S$- and $D$-wave
functions expand continuously. At the critical temperature of
deconfinement $T_c$, while the peak values of the wave functions
drop down a little, their behavior is similar to the one in vacuum.
This means that the $J/\psi$ meson is still a bound state at the
deconfinement phase transition where the light mesons start to melt
in the hot medium, and therefore the observed $J/\psi$s in the final
state of heavy ion collisions can signal the QGP formation in the
early stage. However, with further increasing temperature, the wave
functions expand rapidly around $T_d=1.35 T_c$. By calculating the
average distance between the $c$ and $\bar c$,
\begin{equation}
\label{r} \langle r\rangle (T)={\int dr r^3\left|\psi(r,T)\right|^2
\over \int dr r^2\left|\psi(r,T)\right|^2},
\end{equation}
we found that at $T_d$ the average size becomes infinite, see
Fig.\ref{fig2}, which indicates the $J/\psi$ dissociation.
Therefore, $T_d$ is called the relativistic dissociation
temperature.

The other quantity which can be used to characterize the quarkonium
dissociation is the heavy quark binding energy. For the Dirac
equation or equivalently the Schr\"odinger equations (\ref{wave3})
and (\ref{wave1}), the usual binding energy defined as $\epsilon =
V(\infty)+2m_c-m_m$ is no longer valid to describe the quarkonium
dissociation. In the limit of $r\to\infty$, the four equations for
the spin singlet and triplet degenerate and the asymptotic equation
is simplifies as
\begin{equation}
\label{asym} \left(-{d^2\over
dr^2}+2m_wV(\infty)+V^2(\infty)\right)u=b^2u
\end{equation}
with $V(\infty)=B(\infty)$, and the energy for the scattering state
is
\begin{equation}
\label{scatter}
2m_wV(\infty)+V^2(\infty)=b^2,
\end{equation}
which leads to the binding energy for the $c\bar c$ bound state
\begin{equation}
\epsilon(T) = V(\infty,T)+\sqrt{V^2(\infty,T)+4m_c^2}-m_m.
\end{equation}

\begin{figure}[!h]
\begin{center}
\includegraphics[width=0.35\textwidth]{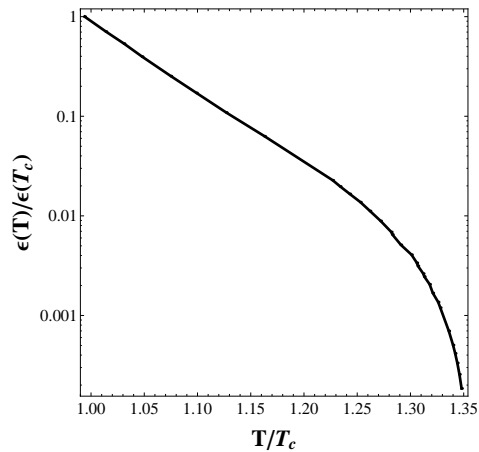}
\caption{ The binding energy of the $J/\psi$ meson in the
quark-gluon plasma and in the limit of potential $V=F$.  The
temperature and binding energy are scaled by their corresponding
values at the critical point of the deconfinement. } \label{fig3}
\end{center}
\end{figure}
Fig.\ref{fig3} shows the temperature dependence of the binding
energy in the QGP phase and in the limit of potential $V=F$. It
drops down monotonously with increasing temperature and reaches zero
at $T_d$. The result is consistent with the calculation of the
charmonium average size. The infinite size and zero binding energy
of the $c\bar c$ system define the unique dissociation temperature.
In comparison with the non-relativistic calculation, the $J/\psi$
dissociation temperature increases from $1.26 T_c$ to $1.35 T_c$,
the relativistic correction is $7\%$. We also calculated the
$J/\psi$ wave functions at finite temperature in the other limit of
potential $V=U$ and found that the dissociation temperature $T_d$
goes up from the non-relativistic value $2.1 T_c$ to $2.38 T_c$. The
relativistic correction is $13\%$.

\begin{figure}[!h]
\begin{center}
\includegraphics[width=0.35\textwidth]{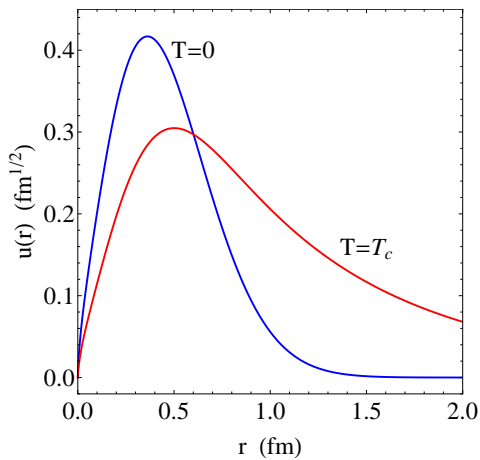}
\caption{ The wave function $u(r)=r \Psi(r)$ for the $\chi_0$ meson
at zero temperature and critical temperature of deconfinement $T_c$
in the limit of quark potential $V=F$. } \label{fig4}
\end{center}
\end{figure}
The wave function for the excited state $\chi_0$ is shown in
Fig.\ref{fig4} in the limit of potential $V=F$. Considering the fact
that the radial function $\Psi$ is divergent at the origin\cite{crater2}, we plot $u=r\Psi$ directly from the Schr\"odinger
equation (\ref{wave3}). In comparison with the ground state
$J/\psi$, $\chi_0$ wave function distributes in a wider region, the
average size is larger and the binding energy is smaller. The
corresponding dissociation temperature defined by $\langle r\rangle
(T_d)=\infty$ and $\epsilon(T_d)=0$ is around the critical
temperature $T_c$ and a little bit higher than the non-relativistic
value. It is easy to understand that the relativistic correction for
the excited states should be smaller than the one for the ground
state.

In summary, we calculated the charmonium wave functions at finite
temperature by solving the covariant relativistic Schr\"odinger
equations for the $c\bar c$ spin singlet and triplet states. The
relativistic effect makes the quark potential well more deep, and
charmonia can survive in a more hot medium. By
considering the two limits of the central potential, the
relativistic correction to the $J/\psi$ dissociation temperature in
the QGP phase is in between $7\%$ and $13\%$.

\vspace{0.2cm}
\appendix {\bf Acknowledgement:} The work is supported by the NSFC (Grant Nos. 10975084 and 11079024),
RFDP (Grant No.20100002110080 ) and MOST (Grant No.2013CB922000).

\bibliographystyle{unsrt}

\end{document}